\documentclass[journal]{IEEEtran}

\usepackage{url}

\usepackage{cite}
\usepackage{hyperref}
\usepackage{nameref}
\usepackage{array}

\usepackage[american]{babel}
\usepackage{graphicx}
\usepackage{subcaption}
\usepackage{caption}
\usepackage{multirow}

\usepackage{siunitx}
\DeclareSIUnit{\molar}{M}

\usepackage[dvipsnames]{xcolor}
\usepackage{pgfplots}
\usepackage{amsmath,amssymb,amsthm,amsfonts}

\usepackage[nomessages]{fp}
\usepackage{booktabs}
\graphicspath{{./src/}}

\newcommand{\gettitle}{Inferring respiratory and circulatory parameters from electrical impedance tomography with deep recurrent models}

\newcolumntype{L}[1]{>{\raggedright\let\newline\\\arraybackslash\hspace{0pt}}m{#1}}
\newcolumntype{C}[1]{>{\centering\let\newline\\\arraybackslash\hspace{0pt}}m{#1}}
\newcolumntype{R}[1]{>{\raggedleft\let\newline\\\arraybackslash\hspace{0pt}}m{#1}}

\hypersetup{
    pdftitle={\gettitle},
    pdfauthor={Nils Strodthoff and Claas Strodthoff and Tobias Becher and Norbert Weiler and Inez Frerichs},
    pdfkeywords={electrical impedance tomography} {bioimpedance} {regression} {deep neural networks} {spirometry} {respiratory system mechanics} {arterial blood pressure}
    bookmarksopen=false,
    bookmarksnumbered=true
}

\begin{document}

\title{\gettitle}

\author{Nils Strodthoff$^{*}$\thanks{This work was supported by the Bundesministerium f\"ur Bildung und Forschung through the BIFOLD - Berlin Institute for the Foundations of Learning and Data (ref. 01IS18025A and ref 01IS18037A)
and by the Faculty of Medicine of the Christian Albrechts University in Kiel, Germany (intramural research funding of research rotation positions).
Nils Strodthoff is with Fraunhofer Heinrich Hertz Institute, Berlin, Germany, e-mail: nils.strodthoff@hhi.fraunhofer.de.
Claas Strodthoff, Tobias Becher, Norbert Weiler and In\'{e}z Frerichs are with the Department of Anesthesiology and Intensive Care Medicine, University Medical Center Schleswig-Holstein, Campus Kiel, Kiel,
Germany, e-mail: \{claas.strodthoff,tobias.becher,norbert.weiler,inez.frerichs@uksh.de\}. Corresponding author is marked by $^{*}$},%~\IEEEmembership{Member,~IEEE,}
\, Claas Strodthoff, Tobias Becher, Norbert Weiler and In\'{e}z Frerichs}

\bstctlcite{BSTcontrol}

\maketitle
\begin{abstract}
Electrical impedance tomography (EIT) is a noninvasive imaging modality that allows a continuous assessment of changes in regional bioimpedance of different organs. One of its most common biomedical applications is monitoring regional ventilation distribution in critically ill patients treated in intensive care units. In this work, we put forward a proof-of-principle study that demonstrates how one can reconstruct synchronously measured respiratory or circulatory parameters from the EIT image sequence using a deep learning model trained in an end-to-end fashion. We demonstrate that one can accurately infer absolute volume, absolute flow, normalized airway pressure and within certain limitations even the normalized arterial blood pressure from the EIT signal alone, in a way that generalizes to unseen patients without prior calibration. As an outlook with direct clinical relevance, we furthermore demonstrate the feasibility of reconstructing the absolute transpulmonary pressure from a combination of EIT and absolute airway pressure, as a way to potentially replace the invasive measurement of esophageal pressure. With these results, we hope to stimulate further studies building on the framework put forward in this work. %As the mentioned parameters are routinely monitored in a clinical setup, these tasks mainly serve as demonstration which information is hidden in the EIT signal and to which degree these can be reconstructed.
\end{abstract}

\IEEEpeerreviewmaketitle
%3-10

\section{Introduction}
\label{sec:intro}
Electrical impedance tomography (EIT) is an imaging modality with a wide range of applications. In the medical field it has primarily been used for functional lung imaging, see \cite{Frerichs2016} for a dedicated review. Here, its advantages are the high rate of image acquisition and the fact that no ionizing radiation is used.
In clinical practice, 16 to 32 electrodes are attached along the thorax circumference. Small currents are injected between changing pairs of electrodes and the resulting voltages are measured between the remaining electrodes. The final step is image reconstruction, where these voltages are converted into cross sectional images by solving the underlying inverse problem\cite{Morucci1996, Grychtol2014}.

With these reconstructed images, spatial patterns of regional aeration changes can be visually inspected, oftentimes by considering the change with respect to the last end of expiration. Different numerical parameters have been found clinically useful, for example measures of variability along the left-right or ventral-dorsal axis (center of ventilation) and measures of ventilation inhomogeneity (global inhomogeneity index; coefficient of variation)\cite{Frerichs2019}. It has been shown that the impedance signal summed over the whole thoracic cross section correlates well with the inspired volume\cite{Ngo2016}.
A number of more sophisticated approaches have been applied that make use of the high frame rate to describe the breath-wise time dependence of aeration of different areas of the lung with exponential or polynomial fits.
Details about these and other measures can be found in \cite{Frerichs2016}.
All of these measures have in common that they are handcrafted and focus on singular arbitrary aspects of the reconstructed EIT image series. This makes interpretation intuitive but obviously reduces the information gained from EIT very crudely if used exclusively. Most of the acquired information is lost when relying on these approaches, which is where this study comes into play.

Mechanically ventilated patients in the intensive care setting are monitored by an increasing number of devices.
Spirometry, the measurement of pressure, volume and flow, is usually integrated in the ventilator and is necessary for selecting and controlling standard ventilator settings.
Transpulmonary pressure is the part of airway pressure that acts upon the lung in contrast to the part that acts upon the chest wall. It is used to adapt ventilator settings to the variable mechanics of the patient’s lungs and chest wall with the aim of minimizing ventilator-induced lung injuries\cite{pes-usage}.

The common way to assess transpulmonary pressure is to use a specialized catheter to measure esophageal pressure which is a good surrogate for the intrapleural pressure. Transpulmonary pressure is then calculated by subtracting esophageal pressure from airway pressure, i.e.\ $p_\text{tp}=p_\text{aw}-p_\text{es}$.
The arterial blood pressure is most commonly measured with an arterial cannula placed in the radial or femoral artery. Systolic, diastolic and mean arterial blood pressures, as well as the shape of the pressure curve offer information about the circulatory system and volume status.
Apart from transpulmonary pressure and EIT, taking these measurements is standard care in intensive care units.

For large-scale biomedical time series data, such as electrocardiography (ECG) data, deep learning methods are the predominant approach \cite{Hannun2019,Ribeiro2020}. While regression tasks are less well studied in this respect, some interesting applications like accurately inferring a patient's age from their ECG have been published\cite{Attia2019,Strodthoff:2020Deep}. One has to distinguish such regression tasks with a single target per sequence from regression tasks that aim to predict a numerical output for all/specific time steps in the sequence. The best-known example for these kinds of problems are time series forecasting tasks\cite{benidis2020neural}, which are after time series classification the most common time series analysis tasks and aim to predict the progression of the time series a few time steps into the future. Only very recently \cite{tan2020time}, another time series annotation task was formalized under the name of \textit{time series regression}, which refers to the task of inferring one or multiple continuous target values per time step, which is precisely the setting investigated in this work. However, a complication arises from the fact that we are not directly trying to identify different time series but are rather dealing with a sequence of images. In this work, we approach this challenge by using a convolutional neural network as feature extractor that transforms the image sequence into a latent representation that is then subsequently processed with a standard recurrent neural network.

\section{Materials and Methods}
\label{sec:methods}
\subsection{Dataset and Tasks}
The data used in this study stems from a study performed to establish an algorithm to optimize ventilation in intensive care patients\cite{becherunpublished}.
20 patients were monitored with EIT (Pulmovista 500, Dr\"{a}ger, L\"{u}beck, Germany) over the course of over 2 hours while different ventilation maneuvers were performed (expiration/inspiration hold, low-flow-loop, step changes of different ventilator settings, different ventilator modes). Patients with corrupted or insufficient aligned data were excluded. All patients had acute respiratory distress syndrome (ARDS) (11 male, median age 71 years). Along with the EIT signal, spirometry (airway pressure $p_\text{aw}$, flow $F$ and volume $V$) was recorded. Many measurements also included arterial blood pressure $p_\text{ab}$, central venous blood pressure and esophageal pressure $p_\text{es}$.

The data used in this study was recorded using three different devices: EIT recorded directly by the EIT device; ventilator data exported via the Medibus protocol ($p_\text{aw}$,$F$,$V$) and data recorded using the Datex Ohmeda S/5 monitor ($p_\text{aw}$,$F$,$V$,$p_\text{ab}$,$p_\text{es}$). This resulted in variable delays between the EIT signal, spirometry and other pressure measurements. We corrected for these delays with a mixture of automatic and manual methods. The small delays between the EIT and the spirometry measurements were aligned by maximizing the cross-correlation of the EIT sum signal and the respiratory volume measurement. The airway pressure measurements recorded by the Datex S/5 monitor were aligned based on the airway pressure signal that was also included in the spirometry measurements. To this end, the signals were roughly aligned using the available metadata and then fine-aligned using the cross-correlation between both signals. During this step, all segments were visually checked for correspondence. The alignment process of the signals of three involved devices represented a major technical challenge in this study. The additional alignment step between Medibus and Datex channels, where only visually matching segments were accepted, leads to a considerable reduction of the available aligned training data for the case of arterial blood pressure and esophageal pressure. In addition to the delays of technical origin, there is a variable physiological delay between cardiac action and the resulting arterial blood pressure wave which depends on numerous factors including state of the cardiovascular system and location of measurement. The position of the arterial cannula was variable and we decided against correcting this delay, which potentially leads to a undefined phase difference between the EIT signal and the arterial blood pressure signal.

All data was downsampled to 10 Hz as a compromise between a sufficiently high temporal resolution to capture for example the arterial blood pressure signal and shorter sequence lengths for a targeted time interval of around 12 seconds, which facilitate the training process. From the synchronized signals we took random crops of 12.8 seconds length corresponding to 128 consecutive frames of reconstructed EIT images at a resolution $32\times 32$. As the global normalization of the EIT signal does not convey any information, we standard-scaled each EIT input sequence individually. The target data was in most cases only standard-scaled by a global channel-wise factor and an additive constant for numerical convenience. For some prediction task discussed below, we only aimed to predict a normalized output, in which case the target data was standard-scaled using batch statistics.

As typical for machine learning approaches, we aimed to make statements about the generalization performance applied to unseen data. Here, we considered two different scenarios that have consequences for the way the data is split into training and test set: In the intra-patient scenario, we investigated the predictive performance of the algorithm when applied to unseen recordings for patients it was trained on. For the intra-patient split the last three recordings of every patient were put into the test set (which is approximately 20\% of the records). The more challenging inter-patient scenario investigates the generalization to patients that were not seen during training. For the inter-patient split, 10\% of the patients were assigned to the test set.

The considered tasks are as follows:
\begin{enumerate}
    \item predicting \textbf{absolute volume} $V$ from the reconstructed EIT image sequence
    \item predicting \textbf{absolute flow} $F$ from the reconstructed EIT image sequence
    \item predicting \textbf{normalized airway pressure} $\hat{p}_\text{aw}$ from the reconstructed EIT image sequence
    \item predicting \textbf{normalized arterial blood pressure} $\hat{p}_\text{ab}$ from the reconstructed EIT image sequence
    \item predicting \textbf{absolute transpulmonary pressure} $p_\text{tp}$ from the reconstructed EIT image sequence and the airway pressure $p_\text{aw}$
\end{enumerate}
Besides the root mean squared error (RMSE) and the dynamic time warping similarity (DTW) \cite{sakoe1978dynamic}, which are common metrics for evaluating regression/forecasting problems, we evaluate the results visually using the following categories and criteria:
\begin{description}
    \item[+] Frequency and amplitude are in good accordance with the target signal.
    \item[o] The predicted signal has the correct frequency/general shape but has the wrong amplitude.
    \item[-] None of the above categories apply.
\end{description}
Examples for the first two categories for each of the tasks under consideration can be seen in the Results section.

\subsection{Inferring Respiratory and Circulatory Parameters from EIT Data}
\begin{figure}[ht]
    \centering
    \includegraphics[width=\columnwidth]{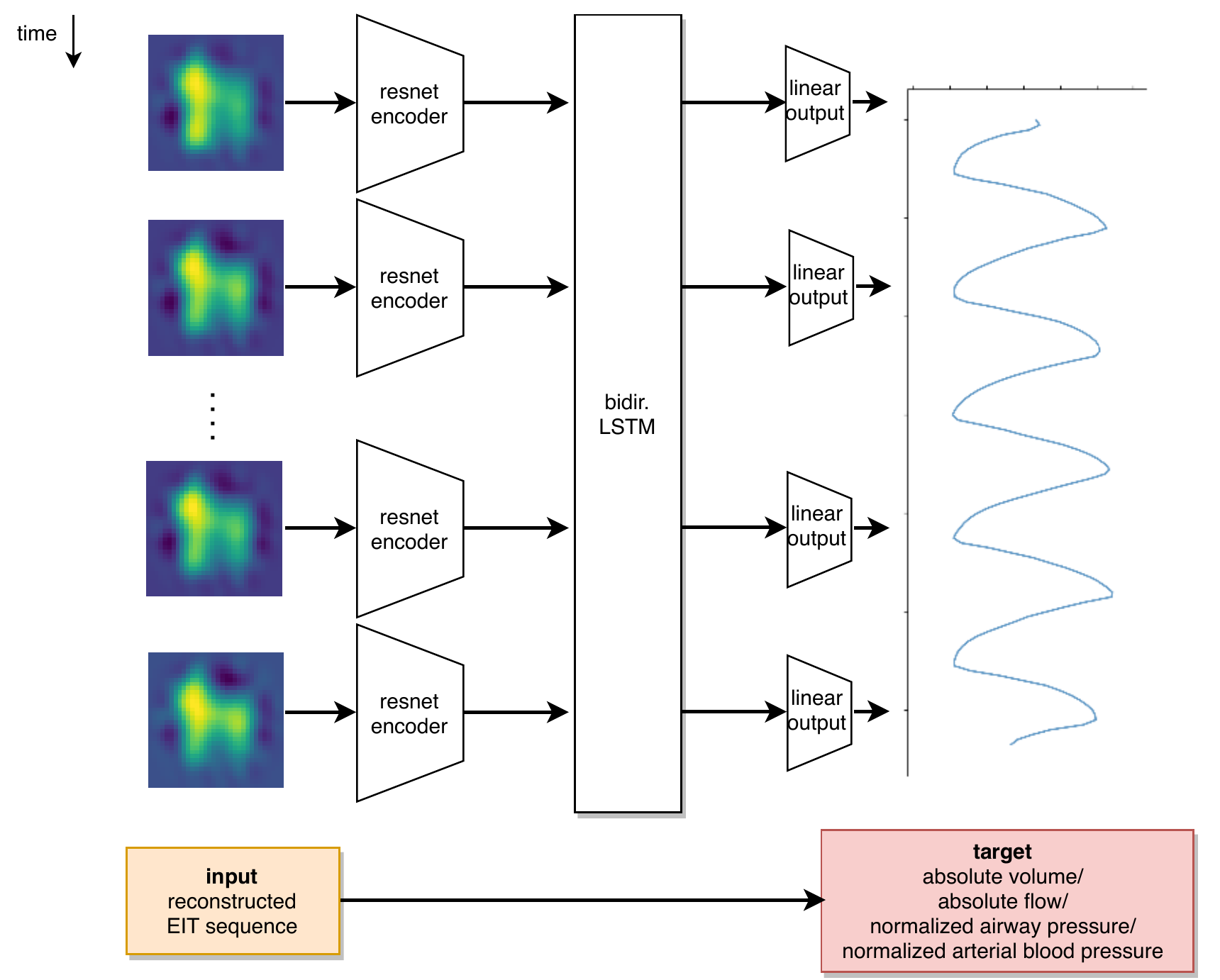}
    \caption{Schematic representation of the bidirectional LSTM model that uses a two-dimensional convolutional neural network as feature extractor.}
    \label{fig:flowchart}
\end{figure}

\begin{table*}[ht]
    \centering
    \caption[]{Prediction performance for spirometry and arterial blood pressure. * denotes shifted RMSE rather than RMSE in the case of arterial blood pressure.}
    \begin{tabular}{l|l|r|r|r|rrr}
        \toprule
        & & target &  & & \multicolumn{3}{c}{visual rating} \\
        split & task & mean($\pm$SD) & RMSE & DTW & + & o & - \\
    \midrule
        \multirow{4}{*}{intra-patient} & absolute volume $V$ & 180($\pm$189) [ml] & 53.60 & 1.864 & 68\% & 32\% & 0\% \\
        & absolute flow $F$ & 0.207($\pm$20.6) [l/s] & 5.963 & 2.383 & 55\% & 45\% & 0\% \\
        & normalized airway pressure $\hat{p}_\text{aw}$ & 0($\pm$1) & 0.2499 & 2.129 & 82\% & 18\% & 0\% \\
        & normalized arterial blood pressure $\hat{p}_\text{ab}$ & 0($\pm$1) & 0.2848* & 4.412 & 69\% & 24\% & 7\% \\
    \midrule
        \multirow{4}{*}{inter-patient} & absolute volume $V$ & 180($\pm$189) [ml] & 57.54 & 1.376 & 70\% & 30\% & 0\% \\
        & absolute flow $F$& 0.207($\pm$20.6) [l/s] & 6.013 & 2.159 & 61\% & 39\% & 0\% \\
        & normalized airway pressure $\hat{p}_\text{aw}$ & 0($\pm$1) & 0.2602 & 1.991 & 88\% & 18\% & 0\% \\
        & normalized arterial blood pressure $\hat{p}_\text{ab}$ & 0($\pm$1) & 0.5365* & 5.678 & 29\% & 43\% & 29\% \\\bottomrule
    \end{tabular}
    \label{tab:results}
\end{table*}

Given a sequence of reconstructed EIT images, the task is to infer one or potentially multiple target quantities discussed in the previous section, i.e.\ the task can be framed as a video regression task where the algorithm has to output a prediction for each frame. The architecture we used to solve this task is comprised of a two-dimensional convolutional neural network as feature extractor (in our case a relatively shallow wide resnet with $N=3$ groups and $n_i=3$ layers per group, i.e.\ in total 19 convolutional layers, $n_f=8$ initial feature dimensions and output dimension $n_\text{intermed}=32$), whose frame-wise output was further processed by a single-layer bidirectional LSTM model \cite{hochreiter1997long} with 512 hidden units followed by linear output layer that outputs a prediction for every frame, see Figure~\ref{fig:flowchart} for a graphical representation of the architecture. This architecture is inspired by related successful literature approaches for video analysis \cite{Donahue2015,McLaughlin2016} and has also been used in the context of functional magnetic resonance imaging analysis\cite{Thomas2019}. Alternative approaches e.g.\ based on the ConvLSTM architecture \cite{Shi2015Convlstm} were also explored but lead to slightly worse results and will therefore not be discussed below.
The model was trained in an end-to-end fashion to minimize the $L_1$-distance(s) between the observed target signal(s) and the predicted output(s). In some cases it turned out to be beneficial to jointly optimize for several target quantities using a shared network with multiple output dimensions. For simplicity, we refrained from introducing different relative factors between the loss terms for the different target quantities. We used the AdamW \cite{loshchilov2017decoupled} optimizer and a 1-cycle learning rate schedule \cite{smith2018disciplined}. Models were implemented using Pytorch \cite{PytorchNIPS2019} and fast.ai \cite{howard2018fastai}.

We used three different variants of the above network to predict transpulmonary pressure: 1)~We predicted the transpulmonary pressure from the EIT image sequence as before. 2)~We jointly predicted transpulmonary pressure and airway pressure from the EIT image sequence. 3)~We included the airway pressure as an additional input, which was then processed by a single linear layer with 32 hidden units with ReLU-activation function and then concatenated with the frame-wise output of the convolutional feature extractor. Thus we inferred the transpulmonary pressure from the combination of airway pressure and EIT image sequence.

\section{Results}
\subsection{Spirometry and Arterial Blood Pressure}

\begin{figure*}[tb]
    \centering
    \includegraphics[width=\textwidth]{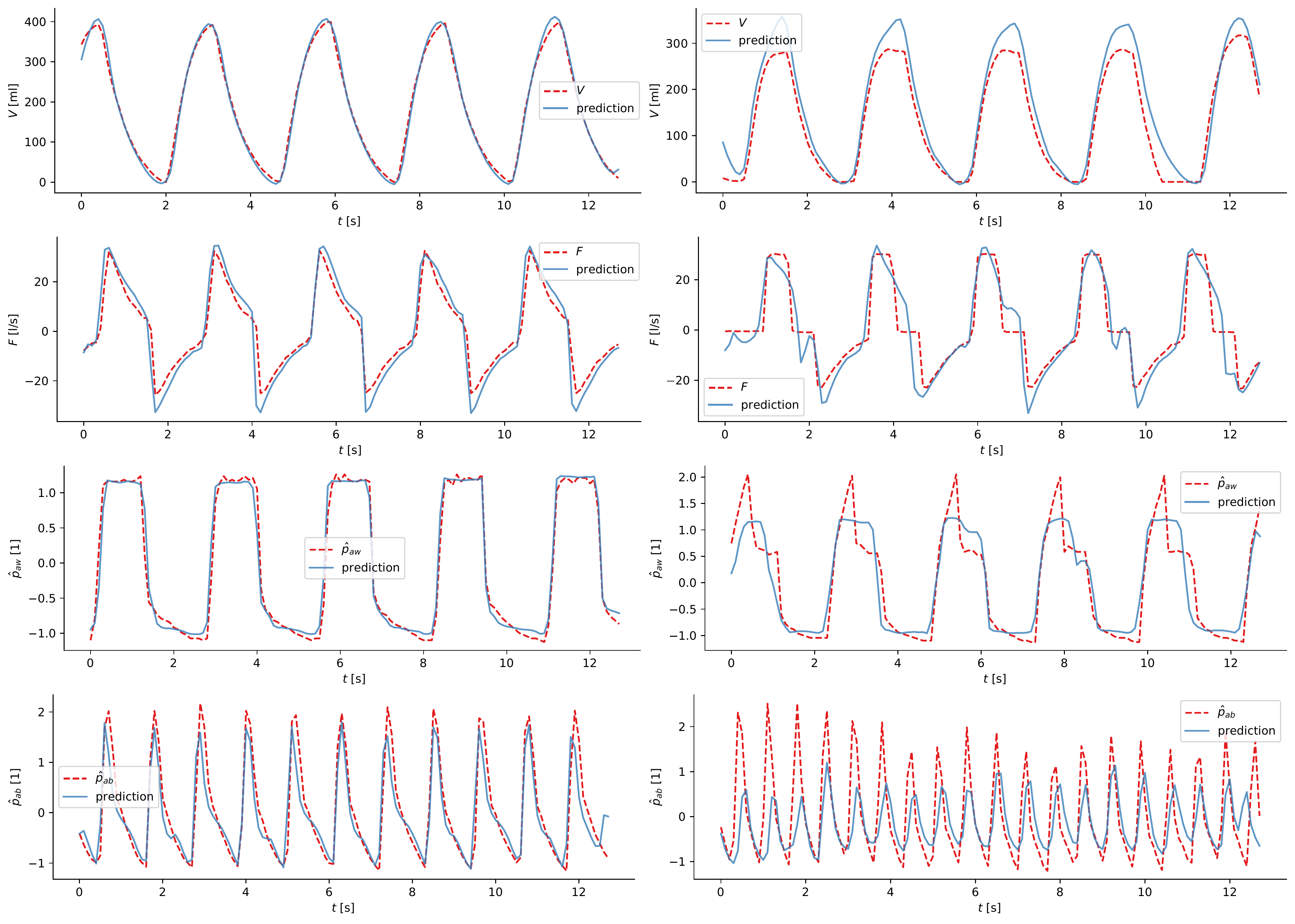}
    \caption{Visual performance summary showing, from top to bottom, exemplary predictions for absolute volume $V$, absolute flow $F$, normalized airway pressure $\hat{p}_\text{aw}$ and normalized arterial blood pressure $\hat{p}_\text{ab}$. The left column shows an example assigned to the best visual rating category ``+'' and the right column an example for the intermediate category ``o''.}
    \label{fig:grid}
\end{figure*}

Due to corrupted and/or insufficiently aligned data, 3 (4 for tasks 4 and 5) patients had to be excluded, resulting in 17 (16) patients.

Table~\ref{tab:results} shows the results for the different tasks both on the record-wise (intra-patient) split and the patient-wise (inter-patient) split based on RMSE and DTW similarity, which are commonly used metrics for regression/forecasting tasks. A corresponding visual performance summary is shown in Figure~\ref{fig:grid}. It was not possible to infer the absolute airway pressure or arterial blood pressure, we therefore restricted to normalized quantities that were standardized using respective batch statistics. For all three spirometric parameters the algorithm shows a solid predictive performance as reflected for example by no samples in the worst category ``-'' and around 70\% ratings in the best category ``+''. The RMSE metric should be set in perspective with the respective standard deviation of the target quantity, which is also listed in Table~\ref{tab:results}. The performance metrics for the spirometric parameters show only minor differences between in the intra-patient setting compared to the more difficult inter-patient setting. The RMSE worsens but interestingly the visual performance metrics and DTW similarity even slightly improve in the inter-patient setting as compared to the intra-patient setting. On the contrary, the arterial blood pressure shows the expected picture, where all performance metrics worsen going from the intra-patient to the inter-patient setting with around 29\% of the segments of the worst visual performance rating in the inter-patient setting, indicating that generalization across patients is difficult for this quantity. Due to the undetermined lag between EIT and arterial blood pressure signal, RMSE is not an appropriate metric in this case as it assumes aligned targets and predictions. We therefore report a shifted RMSE that was obtained by shifting the prediction and the target to the closest maximum in their mutual auto-correlation function. In this context, it is also worth noting, that the training dataset for the arterial blood pressure is considerably smaller than the training datasets for the other three datasets as the arterial blood pressure was captured with a different device than the spirometric observables and required aligned segments.

\subsection{Transpulmonary Pressure}
\begin{table}[t]
    \centering
    \caption[]{Prediction results for the transpulmonary pressure for different combinations of input and output channels using a patient-wise (inter-patient) split. For orientation, the mean($\pm$SD) value of the transpulmonary pressure channel is given by 11.0($\pm$5.67) $\text{cm}\text{H}_2\text{O}$ in this case.}
    \begin{tabular}{l|r|r|rrr}
        \toprule
        & & & \multicolumn{3}{c}{visual rating} \\
     task & RMSE & DTW & + & o & - \\
    \midrule
        EIT $\to$ $p_{tp}$ & 4.847 & 5.898 & 0\% & 96\% & 5\% \\
        EIT $\to$ $p_\text{tp}$, $p_\text{aw}$ & 5.072 & 4.697 & 38\% & 63\% & 0\% \\
        EIT, $p_\text{aw}$ $\to$ $p_\text{tp}$ & 2.841 & 2.631 & 55\% & 45\% & 0\% \\
    \bottomrule
    \end{tabular}
    \label{tab:results-extra}
\end{table}

\begin{figure}[h]
    \centering
    \includegraphics[width=\columnwidth]{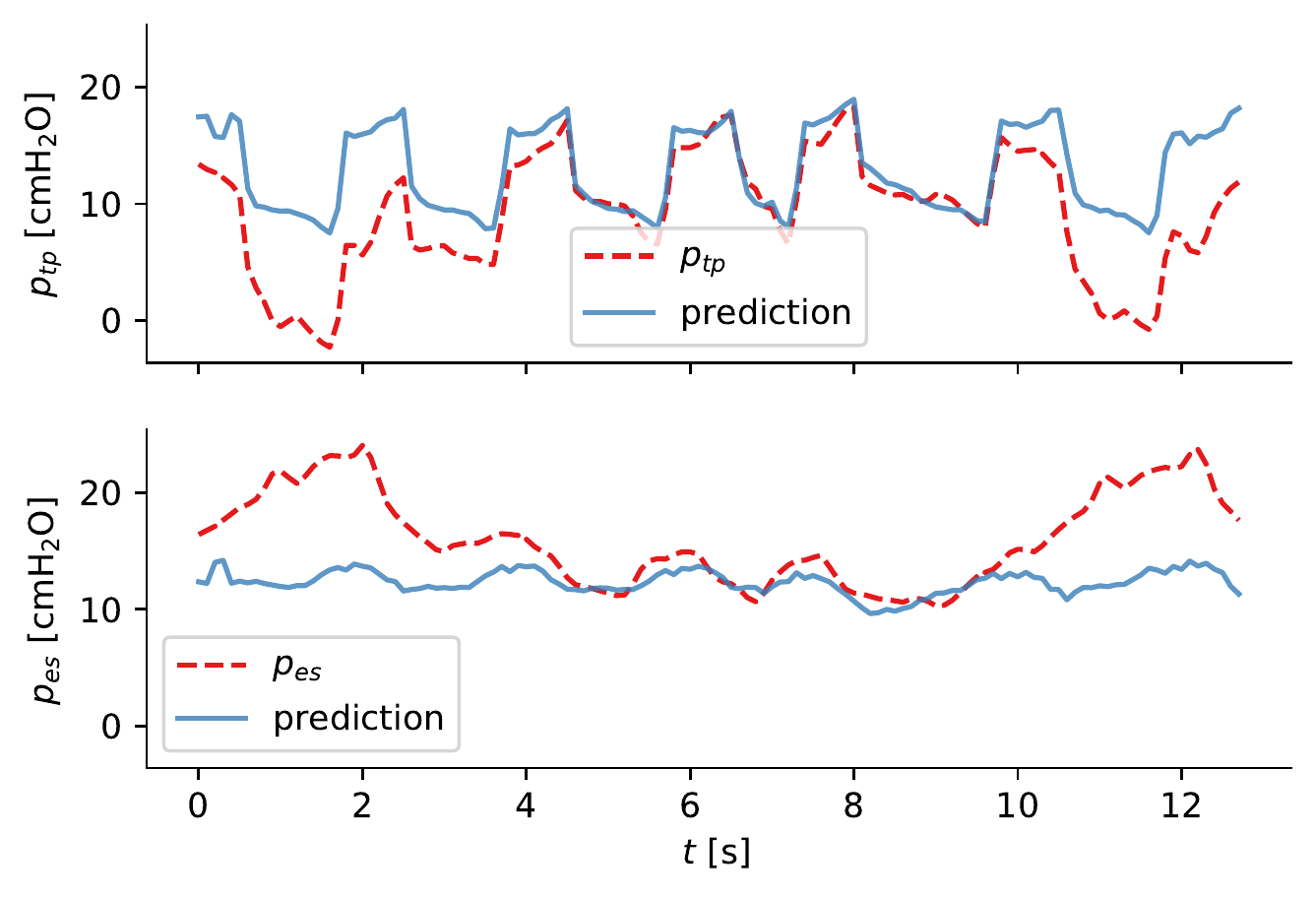}
    \caption{Transpulmonary pressure prediction (top) and corresponding calculated esophageal pressure (bottom).}
    \label{fig:ptp}
\end{figure}

In Table~\ref{tab:results-extra}, we present the results for the additional task of transpulmonary pressure regression on the patient-wise (inter-patient) split. All metrics except RMSE in the second variant show a progression of performance from variants 1 to 3. Interestingly, offering the airway pressure as an extra output seems to improve the model's capacity to predict the general shape (as measured with DTW and visual rating), confirming a known phenomenon\cite{Caruna1996}. The best result by a wide margin is achieved by the model with airway pressure as an extra input. As with the arterial blood pressure, the training dataset is much smaller than the dataset used for the other spirometry-related tasks due to the required extra alignment step between Datex and Medibus channels.
The top panel of Figure~\ref{fig:ptp} shows examples of the predicted and measured transpulmonary pressure as predicted by the best model, which used airway pressure as an extra input. In the bottom panel, we included the corresponding predicted/measured esophageal pressure, obtained by subtracting transpulmonary pressure from the airway pressure.

\section{Discussion}
The results concerning spirometry (Tasks 1-3) are remarkable considering the fact that there are to-date no reliable methods in the literature that are able to perform these tasks and generalize across patients. In fact, the only observable for which there is a real contender for the presented approach is the linear sum model\cite{Ngo2016} for the reconstruction of volume from the EIT signal, which essentially applies a linear regression to the frame-wise sum of the EIT signal. However, the regression coefficients have to be adapted for each patient using a measurement of the actual tidal volume. Therefore it cannot be seen as a generalizing solution in the sense of the intra-patient setting discussed here.

For all patients, most of the recordings were done with pressure-controlled ventilation (constant pressure during inspiration) with only some volume-controlled breaths (constant flow to reach target volume). When the model is confronted with volume-controlled breaths, it performs consistently worse. This is visible in the examples in the right column of Figure~\ref{fig:grid}, where volume, flow and pressure show shapes that are typical for volume-controlled ventilation, as opposed to pressure-controlled ventilation in the left column of Figure~\ref{fig:grid}. This issue would most likely be remedied by a larger and a more diverse dataset.

While spirometry can easily be measured in ventilated patients, this is not true for spontaneously breathing patients.
A possible use of our approach may thus lie in the estimation of absolute tidal volume in non-ventilated patients. The transferability of the approach shown here is uncertain as the physiology of positive pressure ventilation and spontaneous breathing is very different. This transfer to non-ventilated patients would be very useful, because volume calibration for spontaneous breathing requires extra effort\cite{eitlungvolume}. Additionally, this would yield great benefits considering the technical advances in remote patient monitoring with wearable EIT\cite{Frerichs2020}.

The relatively poor capacity of the arterial blood pressure model to generalize to previously unseen patients is probably related to the patient-specific lag between arterial blood pressure and EIT signal due to the variable location of the arterial cannula. In any case, it seems likely that more training data could considerably improve the predictive performance for arterial blood pressure. This is supported by the observation of a considerable gap between the intra-patient and the inter-patient performance, which is not observed for the spirometric observables, suggesting that the network already builds on learned features that generalize across patients. At this point, we stress again that the prediction of the arterial blood pressure mainly serves as demonstration that not only spirometric observables but also circulatory observables can be reconstructed from the EIT signal using the presented approach.

As evident from Figure~\ref{fig:ptp}, the very good result for the transpulmonary pressure regression is not just dominated by the absolute airway pressure that is used as input to the model. The measured esophageal pressure shows slow, positive pressure waves which are an artifact caused by peristaltic contractions of the esophagus, which occur physiologically from time to time. The predicted esophageal/transpulmonary pressures are not subject to this artifact and are thus more reliable. These artifacts were ignored in the visual rating but do (falsely) worsen the RMSE and DTW metrics.
The results on the transpulmonary pressure regression task are very promising, as they might make it possible to integrate the function of pleural pressure measurement into EIT. There would be no need for the costly and potentially harmful invasive esophageal pressure measurement. Furthermore, artifacts caused by esophageal peristalsis could be avoided. The airway pressure needed as extra input is routinely measured by the ventilator anyway. Also in this case, it is to be expected that the model would benefit from more training data to improve generalization.

One of the main limitations of this work is related to the quantity and quality of the data. With only 14 individuals in a typical training set and high variability between individuals, demonstrating the generalization of the presented approach is a challenging task. This is further complicated by variable technical and physiological delays between the signals as well as some signal corruptions. The technical delays between different devices were eliminated as far as possible during preprocessing, but in particular unknown physiological delays as in the case of the arterial blood pressure or the position of the EIT belt remain out of reach. These effects could potentially be compensated for by the algorithm if it was trained on a larger training dataset. The combination of all these issues urges for dedicated studies with synchronized measurements of EIT along with further physiological parameters.

\section{Summary and Conclusions}
The main goal of this study was to demonstrate the feasibility of inferring simultaneously measured spirometric and circulatory parameters from the reconstructed EIT data. To this end, we trained a deep neural network composed of a convolutional feature extractor whose output was fed into a recurrent neural network. This allowed an accurate reconstruction of absolute volume, absolute flow, normalized airway pressure and to certain degree even normalized arterial pressure showing generalization across patients. Furthermore, we demonstrated the same for the transpulmonary pressure using the airway pressure as additional input.

The primary focus of this work was a proof-of-principle demonstration to show how much information is hidden in the EIT signal and how one might extract it. However, as discussed above, there are at least two use-cases, where the presented approach could lead to immediate clinical impact. The first concerns the prediction of spirometric parameters when direct measurement via spirometry is unavailable, which has the potential to make EIT a more versatile and easier tool in these patients. The second concerns the indirect determination of transpulmonary pressure from the EIT signal and airway pressure, which could replace the potentially harmful and sometimes unreliable invasive esophageal pressure measurement.

In the future, one could envision to predict parameters directly from the raw EIT voltage signal without the intermediate EIT image reconstruction step. As the latter involves several filtering processes, skipping this step might even improve the predictive performance. This study represents a strong argument for devising studies where EIT measurements are synchronized with measurements of other circulatory or respiratory parameters, where in the ideal case further progress is fostered by making corresponding datasets publicly accessible for a broader research community.

\bibliographystyle{IEEEtran}

\bibliography{bibfile}

\end{document}